\input{aipcheck}

\documentclass[
    ,final            
  ]
  {aipproc}

\layoutstyle{6x9}

\usepackage{graphicx}
\usepackage{slashed}
\newcommand{\re}{\mathop{\mathrm{Re}}\nolimits}

\newcommand{\notp}{{\slashed{p}}}

\begin{document}

\title{Novel formulations of CKM matrix renormalization}


\author{Bernd A. Kniehl}{
  address={II. Institut f\"ur Theoretische Physik, Universit\"at Hamburg, 
Luruper Chaussee 149, 22761 Hamburg, Germany}
}

\author{Alberto Sirlin}{
  address={Department of Physics, New York University,  
4 Washington Place, New York, NY 10003, USA}
}

\begin{abstract}
We review two recently proposed on-shell schemes for the renormalization of
the Cabibbo-Kobayashi-Maskawa (CKM) quark mixing matrix in the Standard
Model.
One first constructs gauge-independent mass counterterm matrices for the up-
and down-type quarks complying with the hermiticity of the complete mass
matrices.
Diagonalization of the latter then leads to explicit expressions for the CKM
counterterm matrix, which are gauge independent, preserve unitarity, and lead
to renormalized amplitudes that are non-singular in the limit in which any two
quarks become mass degenerate.
One of the schemes also automatically satisfies flavor democracy.
\end{abstract}

\maketitle


\section{Introduction}

\begin{figure}[ht]
\includegraphics[bb=112 626 524 779,width=0.8\textwidth]{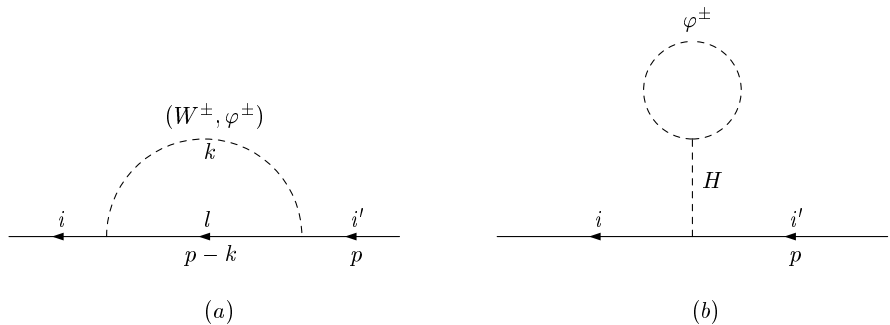}
\caption{\label{fig:one}%
Fermion self-energy diagrams.}
\end{figure}

An important problem associated with the Cabibbo-Kobayashi-Maskawa (CKM)
matrix is its renormalization.
An early discussion \cite{Marciano:1975cn} focused on the renormalization of
ultraviolet (UV) divergences in the two-generation framework.
Since the CKM matrix is one of the fundamental cornerstones of the weak
interactions and, by extension, of the SM, it is important to develop
renormalization schemes that deal with both the finite and divergent
contributions with well-defined renormalization conditions.
Over the last twenty years there have been several papers that address this
basic problem at various levels of generality and complexity.
In this talk I will describe two recently proposed on-shell renormalization
schemes that we think have desirable theoretical properties
\cite{Kniehl:2006bs,Kniehl:2006rc,Kniehl:2009kk}.
The main difficulty arises from external-leg mixing corrections of the type
depicted in Fig.~\ref{fig:one}.

\section{CKM renormalization scheme of Refs.~\cite{Kniehl:2006bs,Kniehl:2006rc}}

Our first proposal is a generalization of Feynman's approach in QED
\cite{Feynman:1949zx}.
Recall that in QED the self-energy contribution to an outgoing fermion is
given by
\begin{eqnarray}
\Delta{\cal M}^{\rm leg}&=&\overline{u}(p)\Sigma(\notp)\frac{1}{\notp-m},
\label{eq:dm}\\
\Sigma(\notp)&=&A+B(\notp-m)+\Sigma_{\rm fin}(\notp),
\end{eqnarray}
where $\Sigma(\notp)$ is the self-energy, $A$ and $B$ are divergent constants,
and $\Sigma_{\rm fin}(\notp)$ is a finite part which is proportional to
$(\notp-m)^2$ in the vicinity of $\notp=m$ and, therefore, vanishes when
inserted in Eq.~(\ref{eq:dm}).
The contribution of $A$ to $\Delta{\cal M}^{\rm leg}$ is singular at $\notp=m$
and is gauge independent, that of $B$ is regular but gauge dependent.
They are called self-mass (sm) and wave-function renormalization (wfr)
contributions.
$A$ is cancelled by the mass counterm $\delta m$, $B$ is combined with proper
vertex diagrams leading to a finite and gauge-independent physical amplitude.
$\Sigma_{\rm fin}(\notp)$ does not contribute to $\Delta{\cal M}^{\rm leg}$.

In the CKM case we encounter not only diagonal terms as in QED but also
off-diagonal contributions generated by the diagram in Fig.~\ref{fig:one}(a).
The self-energy corrections to an external quark leg are now
\begin{equation}
\Delta{\cal M}_{ii^\prime}^{\rm leg}=\overline{u}_i(p)\Sigma_{ii^\prime}(\notp)
\frac{1}{\notp-m_{i^\prime}},
\label{eq:dmii}
\end{equation}
where $i$ denotes the external quark of momentum $p$ and mass $m_i$, and
$i^\prime$ the virtual quark of mass $m_{i^\prime}$.

We focus on the contributions to Eq.~(\ref{eq:dmii}) from Fig.~\ref{fig:one}.
Using a simple algorithm that treats $i$ and $i^\prime$ on an equal footing,
we group the contributions of Fig.~\ref{fig:one} in four classes:
(i) terms with a left factor $(\notp-m_i)$;
(ii) terms with a right factor $(\notp-m_{i^\prime})$;
(iii) terms with a left factor $(\notp-m_i)$ and a right factor
$(\notp-m_{i^\prime})$; and
(iv) constant terms not involving $\notp$.
When inserted in Eq.~(\ref{eq:dmii}), terms of class (iii) vanish.
Terms of classes (i) and (ii) combine to cancel the propagator 
$(\notp-m_{i^\prime})^{-1}$ in both diagonal and off-diagonal contributions.
In analogy with $B$ in QED, they are identified with wfr contributions.
Using the unitarity relations of the CKM elements, one finds that they satisfy
an important property:
all the gauge-dependent and all the UV-divergent wfr contributions to the
$W\to q_i+\overline{q}_j$ amplitude depend only on an overall factor $V_{ij}$
and the external-quark masses $m_i$ and $m_j$, a property shared by the
one-loop proper vertex diagrams.
This implies that, aside from the sm contributions to be discussed later, the
proof of gauge independence and UV finiteness of the one-loop corrections to
the $W\to q_i+\overline{q}_j$ amplitude is the same as in the unmixed,
single-generation case!

In contrast, terms of class (iv) lead to a multiple of
$(\notp-m_{i^\prime})^{-1}$ with a cofactor that involves the chiral
projectors $a_\pm=(1\pm\gamma_5)/2$, but not $\notp$.
In analogy with $A$ in QED, they are gauge independent.
We identify them with the sm contributions.

Next we consider the cancellation of the sm contributions with mass
counterterms
\begin{equation}
\overline{\psi}^Q\left(\delta m^{Q(+)}a_++\delta m^{Q(-)}a_-\right)\psi^Q
\qquad(Q=U,D),
\end{equation}
where $\delta m^{Q(\pm)}$ are non-diagonal matrices subject to the
hermiticity condition
$\delta m^{Q(+)}=\delta m^{Q(-)\dagger}$, which implies
\begin{equation}
\delta m_{i^\prime i}^{U(+)}=\delta m_{ii^\prime}^{U(-)*},\qquad
\delta m_{i^\prime i}^{U(-)}=\delta m_{ii^\prime}^{U(+)*}
\label{eq:mher}
\end{equation}
for $U$ quarks, and similarly for $D$ quarks (for which we use $jj^\prime$).
The UV-divergent sm contributions obey the hermiticity condition, so they can
be cancelled by the $\delta m^{Q(\pm)}$ in all channels.
However, this is not the case for some of the finite parts.
For this reason, we use a specific renormalization prescription:
we adjust the $\delta m^{Q(\pm)}$ to fully cancel the sm terms in all diagonal
channels, as well as the $uc$, $ut$, and $ct$ channels for $U$ quarks and
$sd$, $bd$, and $bs$ channels for $D$ quarks.
This implies that there are residual sm contributions in the reverse $cu$,
$tu$, $tc$, $ds$, $db$, and $sb$ channels, but they are finite, gauge
independent and very small.

An alternative formulation is obtained by diagonalizing the complete mass
matrices $m^Q-\delta m^{Q(+)}a_+-\delta m^{Q(-)}a_-$ by unitary
transformations on the $\psi_R^Q$ and $\psi_L^Q$ fields.
Such transformations induce a CKM counterterm matrix
\begin{equation}
\delta V_{ij}=\sum_{i^\prime\ne i}\frac{m_i^U\delta m_{ii^\prime}^{U(-)}
+\delta m_{ii^\prime}^{U(+)}m_{i^\prime}^U}{\left(m_i^U\right)^2
-\left(m_{i^\prime}^U\right)^2}V_{i^\prime j}
-\sum_{j^\prime\ne j}V_{ij^\prime}
\frac{m_{j^\prime}^D\delta m_{j^\prime j}^{D(-)}
+\delta m_{j^\prime j}^{D(+)}m_j^D}{\left(m_{j^\prime}^D\right)^2
-\left(m_j^D\right)^2}.
\end{equation}
In this basis, in which the complete mass matrices are diagonal, the
off-diagonal sm contributions are cancelled by $\delta V$.
$\delta V$ automatically satisfies the following important properties:
it is gauge independent, preserves unitarity in the sense that both $V$ and
$V-\delta V$ are unitary, and leads to renormalized amplitudes that are
non-singular in the limit $m_{i^\prime}^U\to m_i^U$ or 
$m_{j^\prime}^D\to m_j^D$.

In summary, this approach is based on a novel procedure to separate the
external-leg mixing corrections into gauge-independent sm and gauge-dependent
wfr contributions, and to implement the renormalization of the former with
non-diagonal mass counterterm matrices.
Diagonalization of the complete mass matrices leads to an explicit CKM
counterterm matrix $\delta V$, which is gauge independent, preserves
unitarity, and leads to renormalized amplitudes that are non-singular in the
limit in which any two quarks become mass degenerate.

It has been recently generalized to the case of an extended lepton sector that
includes Dirac and Majorana neutrinos in the framework of the seesaw mechanism
\cite{Almasy:2009kn}.

\section{CKM renormalization scheme of Ref.~\cite{Kniehl:2009kk}}

Our second approach is based on a gauge-independent quark mass counterterm
that is directly expressed in terms of the Lorentz-invariant self-energy
functions.

On covariance grounds, the self-energy $\Sigma_{ii^\prime}(\notp)$ associated
with Fig.~\ref{fig:one} is of the form
\begin{equation}
\Sigma_{ii^\prime}(\notp)=\notp a_-\Sigma_{ii^\prime}^L(p^2)
+\notp a_+\Sigma_{ii^\prime}^R(p^2)
+a_-A_{ii^\prime}^L(p^2)+a_+A_{ii^\prime}^R(p^2).
\label{eq:sig}
\end{equation}
When $i$ is an outgoing $U$ quark, the proposed mass counterterm is
\begin{eqnarray}
\delta m_{ii^\prime}&=&V_{il}V_{li^\prime}^\dagger
\re\left\{a_+\left[\frac{m_{i^\prime}}{2}\tilde{\Sigma}_{ii^\prime}^L(m_i^2)
+\frac{m_i}{2}\tilde{\Sigma}_{ii^\prime}^R(m_i^2)
+\tilde{A}_{ii^\prime}^R(m_i^2)\right]\right.
\nonumber\\
&&{}+\left.a_-\left[\frac{m_i}{2}\tilde{\Sigma}_{ii^\prime}^L(m_{i^\prime}^2)
+\frac{m_{i^\prime}}{2}\tilde{\Sigma}_{ii^\prime}^R(m_{i^\prime}^2)
+\tilde{A}_{ii^\prime}^L(m_{i^\prime}^2)\right]\right\},
\label{eq:delm}
\end{eqnarray}
where $\tilde{\Sigma}_{ii^\prime}^{L,R}(p^2)$ and
$\tilde{A}_{ii^\prime}^{L,R}(p^2)$  are the invariant self-energies with
$V_{il}V_{li^\prime}^\dagger$ factored out.
Explicit one-loop expressions for the invariant functions are given in the
literature \cite{Kniehl:1990mq}.
Inserting these results in Eq.~(\ref{eq:delm}), we get for outgoing $U$
quarks:
\begin{eqnarray}
\delta m_{ii^\prime}^{(+)}&=&\frac{g^2}{32\pi^2}V_{il}V_{li^\prime}^\dagger
\re\left\{m_i\left(1+\frac{m_i^2}{2m_W^2}\Delta\right)
-\frac{m_{i^\prime}m_l^2}{2m_W^2}[3\Delta+I(m_i^2,m_l)
\right.
\nonumber\\
&&{}+J(m_i^2,m_l)]
+\left.m_{i^\prime}\left(1+\frac{m_i^2}{2m_W^2}\right)
[I(m_i^2,m_l)-J(m_i^2,m_l)]
\right\},
\label{eq:delmp}
\end{eqnarray}
where $g$ is the SU(2) gauge coupling, 
$\Delta=1/(n-4)+[\gamma_E-\ln(4\pi)]/2+\ln(m_W/\mu)$ the UV divergence, $n$
the space-time dimensionality, $\gamma_E$ the Euler-Mascheroni constant,
$\mu$ the 't~Hooft mass scale, and
\begin{equation}
\{I(p^2,m_l);J(p^2,m_l)\}=\int_0^1\mathrm{d}x\,\{1;x\}\ln
\frac{m_l^2x+m_W^2(1-x)-p^2x(1-x)-i\varepsilon}{m_W^2}.
\end{equation}
$\delta m_{ii^\prime}^{(-)}$ is obtained by interchanging
$m_i\leftrightarrow m_{i^\prime}$ between the curly brackets in
Eq.~(\ref{eq:delmp}).
Now there are residual sm contributions $\Delta{\cal M}^{\rm res}$ in all
channels, which are finite, gauge-independent, and non-singular in the limits
$m_{i^\prime}\to m_i$ or $m_{j^\prime}\to m_j$, for $m_i,m_j<m_W$.

The mass counterterms $\delta m_{ii^\prime}^{(\pm)}$, as well as
$\delta m_{jj^\prime}^{(\pm)}$ for $D$ quarks, obey two important properties:
(i) they are gauge independent, and
(ii) they automatically satisfy the hermiticity condition of
Eq.~(\ref{eq:mher}).
This second property implies that they can be applied as they stand to all
diagonal and off-diagonal amplitudes and, in this sense, they are
flavor-democratic since they do not single out particular flavor channels.
Again, diagonalization of the complete mass matrices leads to a
gauge-independent CKM counterterm matrix $\delta V$ that preserves unitarity
and now satisfies another highly desirable theoretical property, namely
flavor democracy.

A comparative analysis of the $W$-boson hadronic widths in various CKM
renormalization schemes, including the ones introduced in
Refs.~\cite{Kniehl:2006bs,Kniehl:2006rc,Kniehl:2009kk}, is presented in
Ref.~\cite{Almasy:2008ep}.

This work was supported in part by DFG Collaborative Research Center No.\
SFB~676 and NSF Grant No.\ PHY--0758032.






\IfFileExists{\jobname.bbl}{}
 {\typeout{}
  \typeout{******************************************}
  \typeout{** Please run "bibtex \jobname" to optain}
  \typeout{** the bibliography and then re-run LaTeX}
  \typeout{** twice to fix the references!}
  \typeout{******************************************}
  \typeout{}
 }



\end{document}